\documentclass[12pt]{article}

\usepackage{epsfig,rotating}   

\newcommand{\pcc}{cm$^{-3}$}

\newcommand{\dedx}{$dE/dx$}

\newcommand{\lappeq}{\mathrel{\rlap{\raise.5ex\hbox{$<$}}{\lower.5ex\hbox{$\sim$}}}}
\newcommand{\gappeq}{\mathrel{\rlap{\raise.5ex\hbox{$>$}}{\lower.5ex\hbox{$\sim$}}}}

\begin{document}           

\pagestyle{empty}

\begin{titlepage}

\begin{center}

EUROPEAN ORGANIZATION FOR NUCLEAR RESEARCH

{\small
\begin{tabbing}
 \=  \hspace{117mm}  \=  \kill 
 \>  \>CERN/SPSC 2000-041 \\
 \>  \>SPSC/P317 Add.2  \\ 
 \>  \>October 17, 2000     \\  
\end{tabbing}  }

\vspace{-30mm}
\begin{figure}[htbp]
  \begin{center}
      \makebox{\epsfig{file=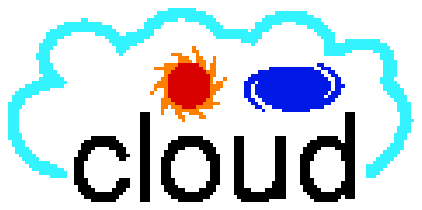,height=18mm} \hspace{0.2mm}}
  \end{center}  
\end{figure}
\vspace{-8mm}

\textbf{CLOUD: AN ATMOSPHERIC RESEARCH FACILITY AT CERN}
\\[2ex] 

 {\footnotesize
 B.\,Fastrup, E.\,Pedersen \\ 
 \emph{University of Aarhus, Institute of Physics and Astronomy, 
 Aarhus, Denmark}  \\[1ex] 
 E.\,Lillestol  \\
 \emph{University of Bergen, Institute of Physics, Bergen,  Norway}  
 \\[1ex]
 M.\,Bosteels, A.\,Gonidec, J.\,Kirkby*, 
 S.\,Mele, P.\,Minginette, B.\,Nicquevert, 
 D.\,Schinzel, W.\,Seidl \\ 
 \emph{CERN, Geneva, Switzerland} \\[1ex] 
 P.\,Grunds\mbox{\o}e, N.\,Marsh, J.\,Polny, H.\,Svensmark  \\
 \emph{Danish Space Research Institute, Copenhagen, Denmark} \\[1ex]  
  Y.\,Viisanen  \\
 \emph{Finnish Meteorological Institute, Helsinki,
 Finland} \\[1ex]
 K.\,Kurvinen, R.\,Orava  \\
 \emph{University of Helsinki, Institute of Physics,  
 Helsinki, Finland}  \\[1ex] 
 K.\,H\"{a}meri, M.\,Kulmala, L.\,Laakso, 
 C.D.\,O'Dowd \\ 
 \emph{University of Helsinki, Lab.\,of Aerosol and
 Environmental Physics, Helsinki, Finland} \\[1ex]
 V.\,Afrosimov, A.\,Basalaev, M.\,Panov  \\ 
 \emph{Ioffe Physical Technical Institute, Dept.\,of Fusion
 Technology, St.\,Petersburg, Russia} \\[1ex]
 A.\,Laaksonen, J.\,Joutsensaari  \\
 \emph{University of Kuopio, Department of Applied Physics, Kuopio,
 Finland}  \\[1ex] 
 V.\,Ermakov, V.\,Makhmutov, O.\,Maksumov, P.\,Pokrevsky,
 Y.\,Stozhkov, N.\,Svirzhevsky  \\
 \emph{Lebedev Physical Institute, Solar and Cosmic Ray Research
 Laboratory, Moscow, Russia} \\[1ex] 
 K.\,Carslaw, Y.\,Yin  \\
 \emph{University of Leeds, School of the Environment, Leeds, United
 Kingdom} \\[1ex]
 T.\,Trautmann  \\
 \emph{University of Mainz, Institute for Atmospheric Physics,
  Mainz, Germany} \\[1ex]
 F.\,Arnold, K.-H.\,Wohlfrom  \\
 \emph{Max-Planck Institute for Nuclear Physics (MPIK), Atmospheric
Physics Division,
 Heidelberg, Germany} \\[1ex]
 D.\,Hagen, J.\,Schmitt, P.\,Whitefield  \\
 \emph{University of Missouri-Rolla, Cloud and Aerosol Sciences
 Laboratory,  Rolla, USA} \\[1ex] 
 K.L.\,Aplin, R.G.\,Harrison  \\
 \emph{University of Reading, Department of Meteorology,
 Reading, United Kingdom} \\[1ex]
 R.\,Bingham, F.\,Close, C.\,Gibbins, A.\,Irving, B.\,Kellett,
M.\,Lockwood \\
 \emph{Rutherford Appleton Laboratory, 
 Space Science \& Particle Physics Depts., Chilton, United
 Kingdom} \\[1ex] 
  J.M.\,M\"{a}kel\"{a} \\
 \emph{Tampere University of Technology, Department of Physics,
Tampere,  Finland} \\[1ex] 
 D.\,Petersen, W.W.\,Szymanski, P.E.\,Wagner, A.\,Vrtala  \\ 
 \emph{University of Vienna, Institute for Experimental Physics,
 Vienna, Austria}  \\[1ex]   }                        
 {\normalsize CLOUD$^\dagger$  Collaboration} \\
\end{center}
\vspace{-5mm}

\vfill

\noindent \rule{60mm}{0.1mm} \\ {\footnotesize
$*$) spokesperson  \\
$\dagger$) Cosmics Leaving OUtdoor Droplets }
\date{}

\end{titlepage}

\pagestyle{plain}     

\pagenumbering{roman}  
\setcounter{page}{2}  
\newpage \tableofcontents 


\newpage

\pagestyle{plain}     
\pagenumbering{arabic}  
\setcounter{page}{1}  

\section{Introduction}  \label{sec_introduction}

\subsection{Overview}  \label{sec_overview}

At its meeting on 6 September 2000, the SPSC suggested that CLOUD
\cite{cloud_proposal,cloud_addendum_1} should be considered as a
facility rather than a single experiment. We welcome this suggestion,
which corresponds with earlier informal discussions that we had with
CERN management. This document therefore places CLOUD in the framework
of a CERN facility for atmospheric research. It also addresses some
questions raised by the SPSC at the same meeting, concerning the use of
a CERN PS beam. 

\subsection{An Atmospheric Research Facility at CERN}
\label{sec_facility}

Clouds are the familiar but complex engines of the world's weather and
climate. Basic questions remain unanswered about the physicochemical
production of the aerosols on which the water droplets or ice crystals
form. Under certain conditions, ionisation of the air by cosmic rays
may play a vital part. Theoretical studies and direct observations of
ion effects from aircraft and balloons give important clues, but
adequate laboratory facilities to investigate the microphysics under
controlled conditions on the ground do not yet exist. 

The experimental concept of CLOUD begins with studies of the various
microphysical processes by which cosmic rays may affect cloud formation
in the atmosphere. The initial programme contains five groups of
experiments dealing with (1) nucleation and growth of aerosols, (2)
formation of cloud droplets, (3) production of condensable vapours, (4)
creation of ice nuclei, and (5) dynamics of stratospheric clouds. This
programme is only indicative, and the actual investigations will
respond to our experimental results and associated theoretical
developments. 

\begin{figure}[htbp]  
  \begin{center} 
      \makebox{\epsfig{file=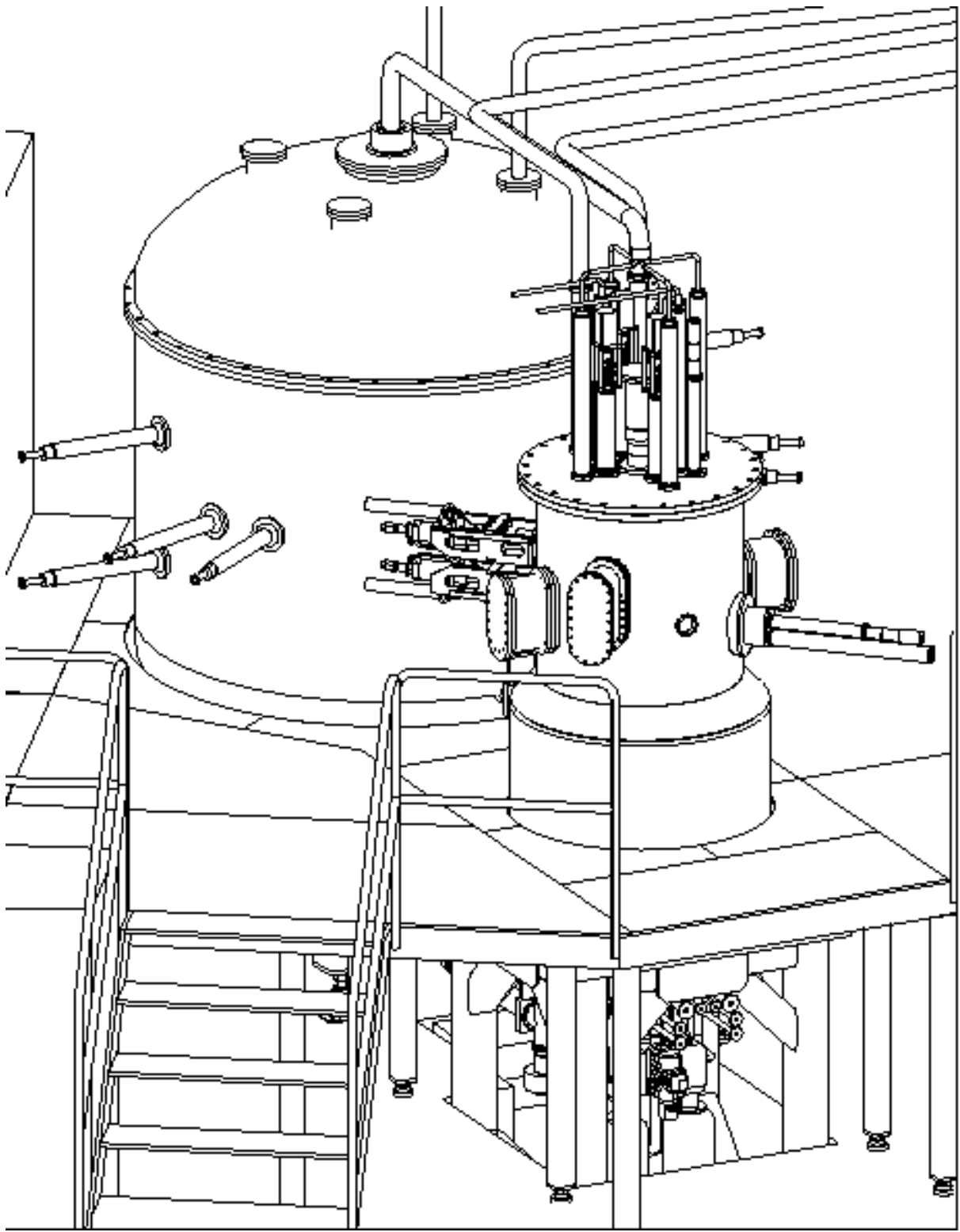,width=160mm}}
  \end{center}
  \caption{Perspective view of the proposed CERN Atmospheric Research
Facility, CLOUD.}
  \label{fig_ cloud_3d_closeup}
\end{figure}

\begin{figure}[htbp]  
  \begin{center} 
      \makebox{\epsfig{file=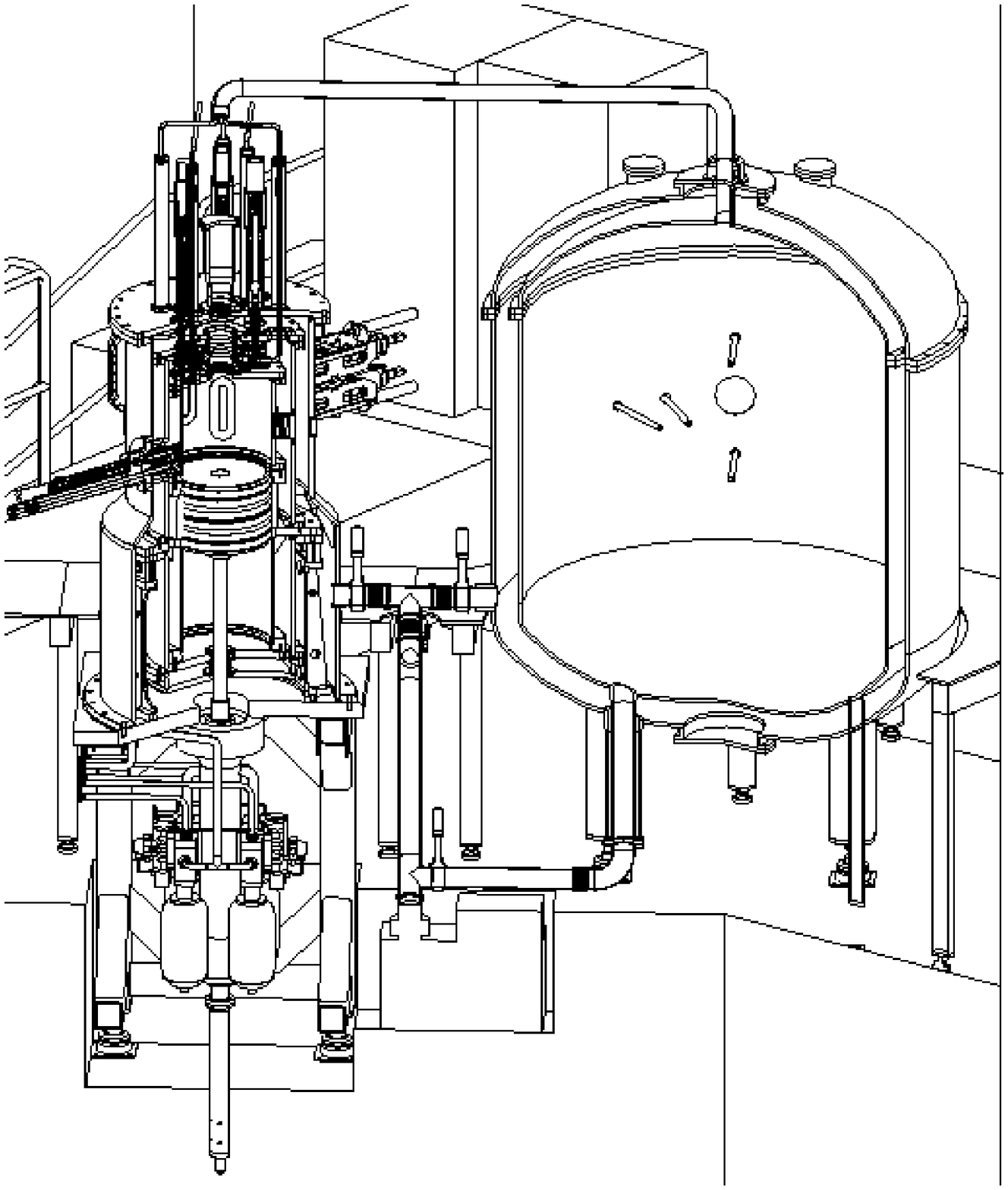,width=160mm}}
  \end{center}
  \caption{Cut-away view of CLOUD.}
  \label{fig_ cloud_3d_closeup_cut}
\end{figure}

For performing the CLOUD investigations we propose (Figs.\,\ref {fig_
cloud_3d_closeup} and \ref {fig_ cloud_3d_closeup_cut}) to combine a
novel expansion cloud chamber with a reactor chamber and associated
experimental systems for gas and particle analysis.  Reasons for
designating CLOUD as CERN's Atmospheric Research Facility are as
follows: 

\begin{itemize}

\item The concept of a facility is appropriate for the comparatively
large and complex experimental programme of CLOUD. 

\item CLOUD is a `general purpose' detector. Flexibility is required
because we do not know what we shall discover or where rapid
experimental and theoretical developments may lead us.   

\item No such facility is presently available to atmospheric scientists
at any other location worldwide. 

\item A time span for the experiments of several years is envisaged. 

\item Our team already includes atmospheric scientists from 10
institutes in 5 countries but we can expect others to propose
experiments when they know of the existence of the facility and its
unrivalled capabilities. 

\end{itemize}

Why at CERN? We consider the use of a CERN particle beam to be crucial.
Carefully-controlled and precisely-delivered particle ionisation at
natural GCR intensities and ionisation densities (corresponding to
minimum ionising particles) is a central part of this experiment.
Experiments by members of our collaboration and others have obtained
useful results with traditional sources of ionisation such as X rays and
radioactive sources. But to make further progress requires a beam from
a particle accelerator.  Only a particle beam can closely duplicate the
characteristics of cosmic rays throughout the atmosphere and deliver a
precisely known ionisation with  the flexibility of intensity, timing,
spatial distribution and penetration range that the facility requires.  

In addition, CLOUD relies heavily on CERN's expertise with
bubble-chambers and cryogenic temperature control (required to within
0.01K stability). We also need assistance from CERN for the technical
integration of what we consider to be a large and complex experiment,
as well as for computing support.  

It is this combination of a particle beam and specialised technical
expertise that makes CERN uniquely suitable for the proposed facility. 
In a broader context, the creation of the Atmospheric Research Facility
will give very positive signals (1) that particle physics has direct
relevance to terrestrial concerns, and (2) that CERN welcomes
cross-disciplinary research proposed by scientists in its member states
who are not particle physicists.

\section{Particle beam}  \label{sec_particle_beam}

\subsection{Beam requirements}
\label{sec_beam_requirements}

The primary task of CLOUD is to investigate how galactic cosmic rays
(GCRs) may influence cloud formation.  To do this we plan to establish
realistic atmospheric conditions inside a cloud chamber and reactor
chamber, and irradiate the chambers with a particle beam to provide a
realistic and  adjustable source of ``cosmic rays''.  We can estimate
the beam intensities required for CLOUD as follows. 

The desired beam intensity is between $1 \times$ and about $10 \times$
the natural GCR flux at a given altitude. This will  allow us to
measure the dependence of any observed effects on ionisation rate and
ion pair  concentration, and to extrapolate reliably across the range
of natural ionisation in the atmosphere.  The highest beam intensities
will help to amplify and expose effects before they are measured at
natural ionisation levels. Beam-off data will be also be recorded,
under conditions with the chamber  clearing fields on and off,
respectively, corresponding to $0.01
\times$ and $1 \times$ the natural GCR ion pair concentrations at ground
level.  The optimum operating  energy of the proposed T11 beam is near
the maximum (3.5~GeV/c) in order to minimise beam particle scattering.

\begin{figure*}[htbp]
  \begin{center}
      \makebox{\epsfig{file=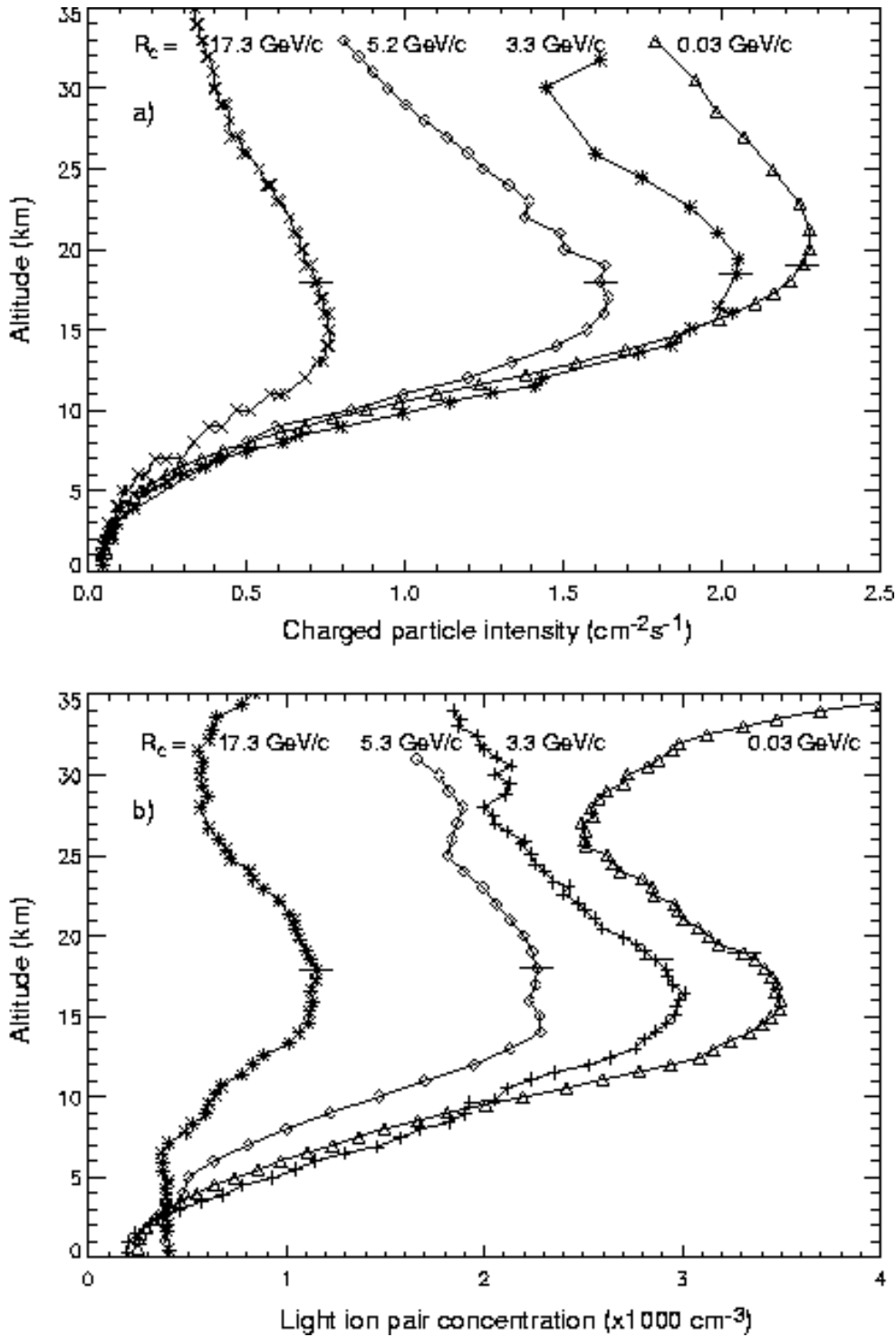,width=130mm}}
  \end{center}
   \vspace{-5mm}
 \caption{ a) The  charged  particle intensity 
  and b) the small ion pair concentration  vs.\,altitude, measured at
several latitudes with cutoff rigidities, $R_c$, as indicated. The data
were recorded by Lebedev Physical Institute \cite{ermakov97} in or near
1990, corresponding to a sunspot maximum (but without solar proton
events), i.e during a cosmic ray
\emph{minimum}. The horizontal  bars  show the typical experimental
statistical errors.}
  \label{fig_cosmics}    
\end{figure*}

CLOUD will measure processes over the full range of tropospheric and
stratospheric conditions.  At ground level, the average GCR intensity
is about 0.02 cm$^{-2}$s$^{-1}$, whereas at altitudes of 15--20 km it
is about a factor 100 larger, varying between about 0.8 and 2.3
cm$^{-2}$s$^{-1}$ depending on geomagnetic latitude 
(Fig.\,\ref{fig_cosmics}a).

The maximum required \emph{time-averaged} beam intensity is therefore
about $10 \times 2 = 20$~cm$^{-2}$s$^{-1}$.  The beam is spread over a
large transverse area of 30 cm $\times$ 30 cm $\simeq$~1000~cm$^2$ in
order to duplicate the quasi-uniform GCR irradiation, as closely as
possible, over  the fiducial volume. This is achieved by defocusing the
quadrupole magnets in the T11 beamline.  The time-averaged maximum beam
intensity is then   $20 \times 1000 =  2 \cdot 10^4$ s$^{-1}$.  If we
assume a 0.5 s beam pulse from the CERN PS every 5 s, then the maximum
beam intensity is $5 \times 2 \cdot 10^4 =  10^5$~/pulse.  This is
comfortably  within the maximum performance of  $7 \times 10^5$~/pulse
for the T11 beamline at 3.5 GeV/c, assuming $2 \times 10^{11}$
protons/pulse on the target. 

The \emph{minimum} beam intensity (apart from beam-off) is $1 \times$
the  natural GCR radiation at ground level.  This is a factor 1000
below the maximum required intensity (a factor 100 for the atmospheric
attenuation and a factor 10 for $1 \times$ the GCR intensity rather than
$10 \times$), i.e.  a time-averaged intensity of 20 s$^{-1}$, or 100
/pulse, assuming 1 pulse every 5 s. 

These beam estimates are summarised in Table
\ref{tab_beam_conditions}.

\begin{table}[htbp]
  \begin{center}
  \caption{Summary of the minimum and maximum beam conditions  for
CLOUD.  One beam pulse per 5 s is assumed, as well as a transverse beam
size of $30 \times 30$ cm$^2$. `GCR' signifies the natural GCR
intensity at the indicated altitude.}
  \label{tab_beam_conditions}
  \vspace{5mm}
  \begin{tabular}{| r c | c c |}
  \hline
 \textbf{Simulated} & \textbf{Simulated} & \textbf{Beam intensity}
 & \textbf{Clearing} \\
  \textbf{GCR intensity} & \textbf{altitude (km)} &
\textbf{(particles/pulse)} 
 & \textbf{field} \\
  \hline
  \hline
  &  &  &  \\[-3ex]
 0.01 $\times$ GCR & 0  & 0 & on \\
    1 $\times$ GCR & 0  & 0 & off \\
    2 $\times$ GCR & 0  & $10^2$ & off \\
   10 $\times$ GCR & 0  & $10^3$ & off \\
   \hline 
 0.01 $\times$ GCR & 10 & 0 & off \\   
    1 $\times$ GCR & 10 & $10^4$ & off \\
   10 $\times$ GCR & 10 & $10^5$ & off \\[0.5ex]
  \hline
  \end{tabular}
  \end{center}
\end{table}

\subsection{Beam simulations}

\begin{figure}[htbp]
  \begin{center}
      \makebox{\epsfig{file=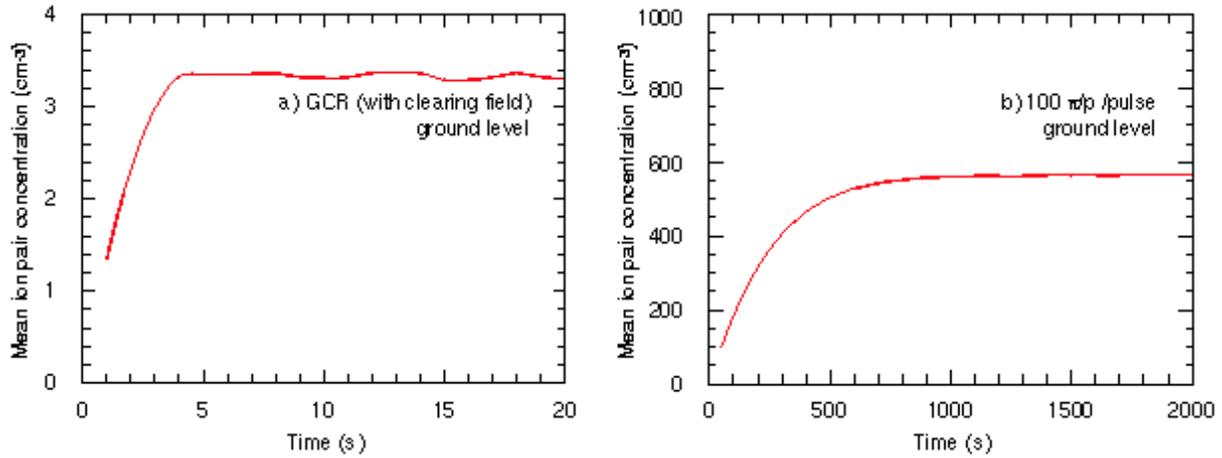,width=160mm}}
  \end{center}
  \caption{Mean ion pair concentration in the cloud chamber vs.\,time
for the ambient GCR flux and a) no beam and clearing field on, and b)
100 particles /pulse ($\times$~1~pulse per 5s) and clearing field off. 
The chamber conditions correspond to aerosol-free air at ground level
(293 K and 101 kPa). The ion pair concentration is assumed to be zero at
t=0.}
  \label{fig_ions_vs_time}    
\end{figure}

\begin{figure}[htbp]
  \begin{center}
      \makebox{\epsfig{file=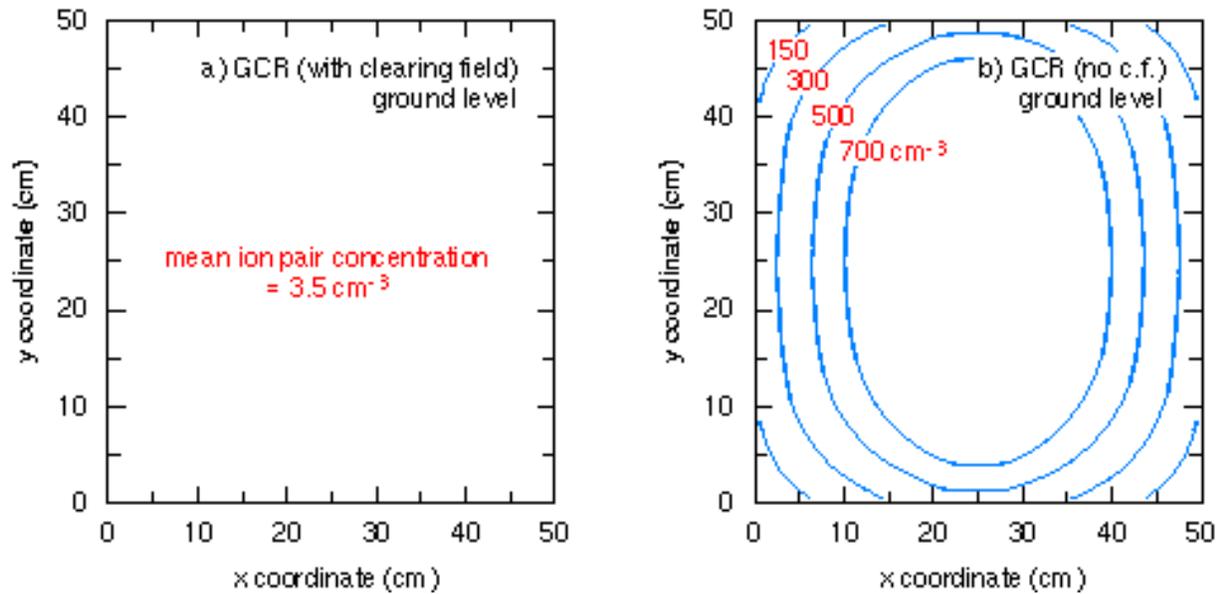,width=160mm}}
  \end{center}
  \caption{The equilibrium ion pair concentrations in the cloud chamber
for no beam and the ambient GCR flux, a) with clearing field, and b)
without clearing field.  The chamber conditions correspond to
aerosol-free air at ground level (293 K and 101 kPa).  The projections
show a central slice through the chamber in the vertical plane, with the
beam axis along $y = 25$~cm.}
  \label{fig_contours_gcr}    
\end{figure}

\begin{figure}[htbp]
  \begin{center}
      \makebox{\epsfig{file=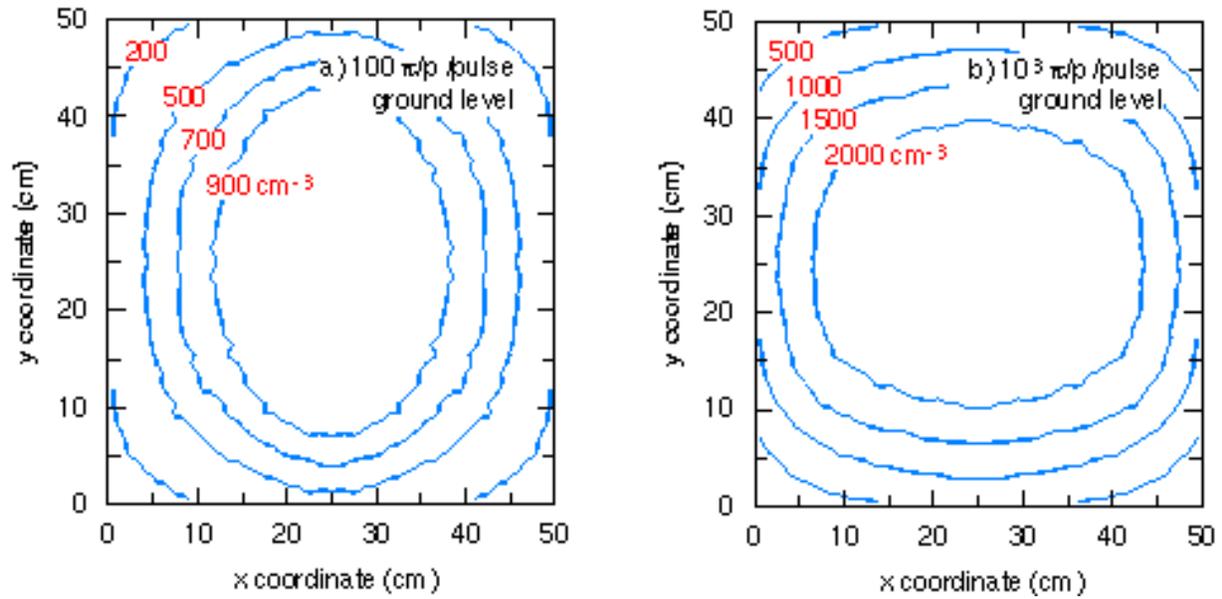,width=160mm}}
  \end{center}
  \caption{The equilibrium ion pair concentrations in the cloud chamber
for the ambient GCR flux and a beam of a) 100 particles /pulse  ($\times
1$~pulse per 5s), and b) $10^3$ particles /pulse. The chamber
conditions correspond to aerosol-free air at ground level (293 K and
101 kPa).}
  \label{fig_contours_0km}    
\end{figure}

\begin{figure}[htbp]
  \begin{center}
      \makebox{\epsfig{file=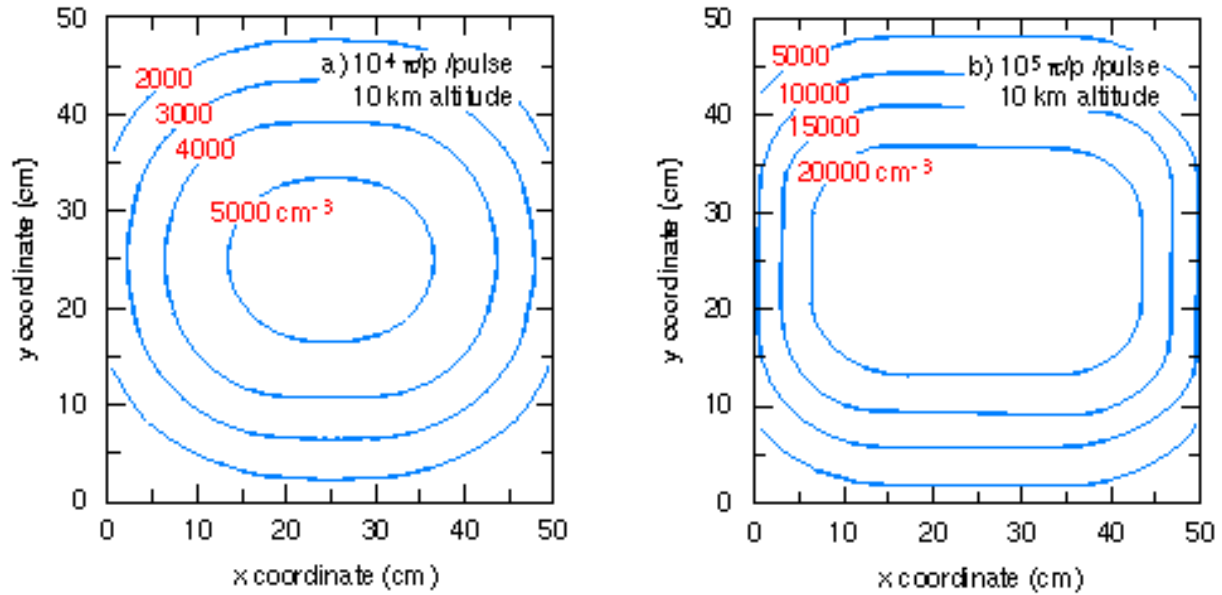,width=160mm}}
  \end{center}
  \caption{The equilibrium ion pair concentrations in the cloud chamber
for a beam of a)~$10^4$ particles /pulse, and b) $10^5$ particles
/pulse. The chamber conditions correspond to aerosol-free air at 10 km
altitude (223 K and 26 kPa). }
  \label{fig_contours_10km}    
\end{figure}

The calculations in \S\ref{sec_beam_requirements} provide useful
estimates of the beam requirements for CLOUD.  However they
under-estimate the actual beam requirements since they take no account
of diffusion losses of ions to the chamber walls or of ion-ion
recombination.    We have therefore performed a 3-dimensional
simulation which includes these effects.  The results are shown in
Figs.\,\ref{fig_ions_vs_time}--\ref{fig_contours_10km} and are
described below.

\paragraph{GCR (no beam) conditions:} The equilibrium ion pair
concentration due to the background GCR flux\footnote{The GCR flux will
be monitored by roof of recuperated plastic scintillation counters over
the CLOUD facility.} is about 500 \pcc\ averaged over the cloud chamber
volume (Fig.\,\ref{fig_contours_gcr}b).  This value is slightly higher
than typical measured values at ground level (Fig.\,\ref{fig_cosmics}b)
since atmospheric aerosols scavenge small ions.  The equilibrium
concentration is reached after about 10 minutes
(Fig.\,\ref{fig_ions_vs_time}b shows a typical time evolution)---which
is characteristic of small ion lifetimes in the atmosphere.   The mean
ion pair concentration can be readily reduced below the natural GCR
level by turning on the clearing field.  With 1 kVm$^{-1}$ electric
field, small ions are cleared from the cloud chamber in 4~s, and this
reduces the mean ion concentration to about 1\% of the atmospheric
values at ground level (Figs.\,\ref{fig_ions_vs_time}a and
\ref{fig_contours_gcr}a).  The clearing field can therefore be used very
effectively to turn off ion-induced processes in the cloud and reactor
chambers, since the ion-mediated effects generally occur on timescales 
that are long (several minutes to hours) compared with the clearing
time.  Note that, depending on their size, charged aerosols are not
swept out by the clearing field due to their low mobilities. 

\paragraph{Ground-level conditions:} Data taken without beam and with
various clearing field settings will provide the lowest ionisation
measurements at ground level conditions.  For higher ionisations at
ground level conditions, the beam is used.  The expected ionisation
concentrations for 100 particles /pulse and $10^3$ particles /pulse are
shown in Figs.\,\ref{fig_contours_0km}a) and b), respectively, assuming
1 pulse /5~s. These beam intensities produce rather modest increases in
the ion pair concentrations since the cross-sectional area of the cloud
chamber illuminated by the beam is about a factor 2.5 less than that
illuminated by GCR.  To obtain the desired factor 10 increase in ion
pair concentration above atmospheric values  requires a beam  intensity
of about $4 \times 10^3$ particles /pulse.   

\paragraph{10 km conditions:}    The expected ionisation concentrations
for $10^4$ particles /pulse and $10^5$ particles /pulse at 10 km
conditions are shown in Figs.\,\ref{fig_contours_10km}a) and b),
respectively. 
 The mean ion pair concentrations at these high altitudes are about
3000~\pcc\ for high geomagnetic latitudes (Fig.\,\ref{fig_cosmics}b).
These beam intensities therefore reasonably well cover the desired
range, although a maximum beam intensity of about $2 \times 10^5$
particles /pulse is required to reach $10 \times$ the natural GCR level
at 10 km altitude.  Finally we note that the ground-level background
GCR in CLOUD decreases in significance for increasing altitude
conditions, and is essentially negligible ($\sim$1\%) at 10~km
conditions.

\subsection{Alternative ionisation sources}

The basic goal of CLOUD is to duplicate  atmospheric and cosmic ray
conditions in the laboratory. Cosmic rays consist mostly of pion and
electron secondaries in the upper troposphere, and decay muons near
ground level.  The primary cosmic rays mostly interact in the
($\sim$2$\lambda$) material above the tropopause.  Therefore,
essentially throughout the troposphere,  charged cosmic rays comprise
minimum ionising particles. 

The requirements of the ionisation source for CLOUD are as follows:
\begin{itemize}
\item Capability to deposit a precisely known quantity of ionisation at 
a precisely known location inside the cloud chamber and reactor chamber.
\item An ionisation density (\dedx) that is characteristic of minimum
ionising particles.
\item Easily adjustable in intensity  over the required range of
1--10$\times$ the natural cosmic ray intensities found in the
troposphere---a factor of about 1000.
\item Ability to traverse the walls and liquid cooling layers of the
cloud chamber and reactor chamber.  This sets a minimum energy for a
particle beam of about 1 GeV/c, taking multiple Coulomb scattering also
into account. 
\item Known timing. This is necessary for the study of fast processes
and also for ice nucleation studies to distinguish between deposition
nucleation and freezing nucleation.
\end{itemize}

Other sources of ionising radiation include ultraviolet (UV)
radiation,  radioactive sources and X ray sources.  UV radiation is
excluded since it induces photochemical reactions among the trace
gases. In the case of radioactive sources,
$\alpha$ emitters are excluded by their high ionisation density and
short range, and $\beta$ emitters are excluded since they cannot be
placed inside the cloud chamber, and their range is insufficient to
penetrate the chamber walls.  

Gamma radioactive sources are impractical; they would need to be
distributed around the outer surfaces of the cloud chamber and reactor
chamber, and even then would provide imprecisely-known and non-uniform
ionisation in the fiducial volumes. The maximum required intensity is
equivalent to 1000 times the ground-level GCR intensity at a distance
of about 1 m from the source.  This represents a significant radiation
hazard and poses handling and safety problems.  Furthermore it is
difficult to adjust the intensity of a gamma source over the required
range, and there are no timing capabilities.  
 X ray sources have similar limitations, with the added problem that,
since the range of X~rays is shorter, large absorption and
non-uniformities result from the detector material.

A careful search of the literature \cite{aplin00b} has uncovered many
little-known laboratory studies of ion-induced effects on aerosol
formation since the 1960's using traditional ionisation sources
(e.g.\,refs.\,\cite{bricard,burke,vohra}). Members of our collaboration
have also studied these processes more recently using X rays and 
$\alpha$ particles from  $^{241}$Am sources \cite{makela}.  Although
some useful results have been obtained, these studies have generally
been unable to characterise the aerosol processes adequately.
 There are two reasons for this: 
\begin{enumerate}
\item Lack of control of the ionisation at near-atmospheric intensities.
\item Non-uniformities of deposited ionisation.  We have shown
\cite{clement} that local variations of ion density (for example, from
an $\alpha$ source) give rise to non-linear aerosol charging effects,
which will directly affect ion-induced aerosol processes.  This makes
it difficult to relate results obtained with such sources to the real
atmosphere.
\end{enumerate} The limitations of traditional sources have hindered
further progress with these experiments.

In contrast, a GeV beam from a particle accelerator ideally matches the
requirements of a well-defined and reproducible ionisation source. It
can deliver a precise quantity of ionisation with a precisely known
spatial distribution, and at an exactly known time. Until now, as far
as we know, no one has used an accelerator beam for such studies. By
combining this ideal ionisation source with an advanced detector of
unrivalled capabilities, CLOUD offers excellent prospects for a
breakthrough in ion-aerosol-cloud research.

\begin{sidewaysfigure}
  \begin{center}
      \makebox{\epsfig{file=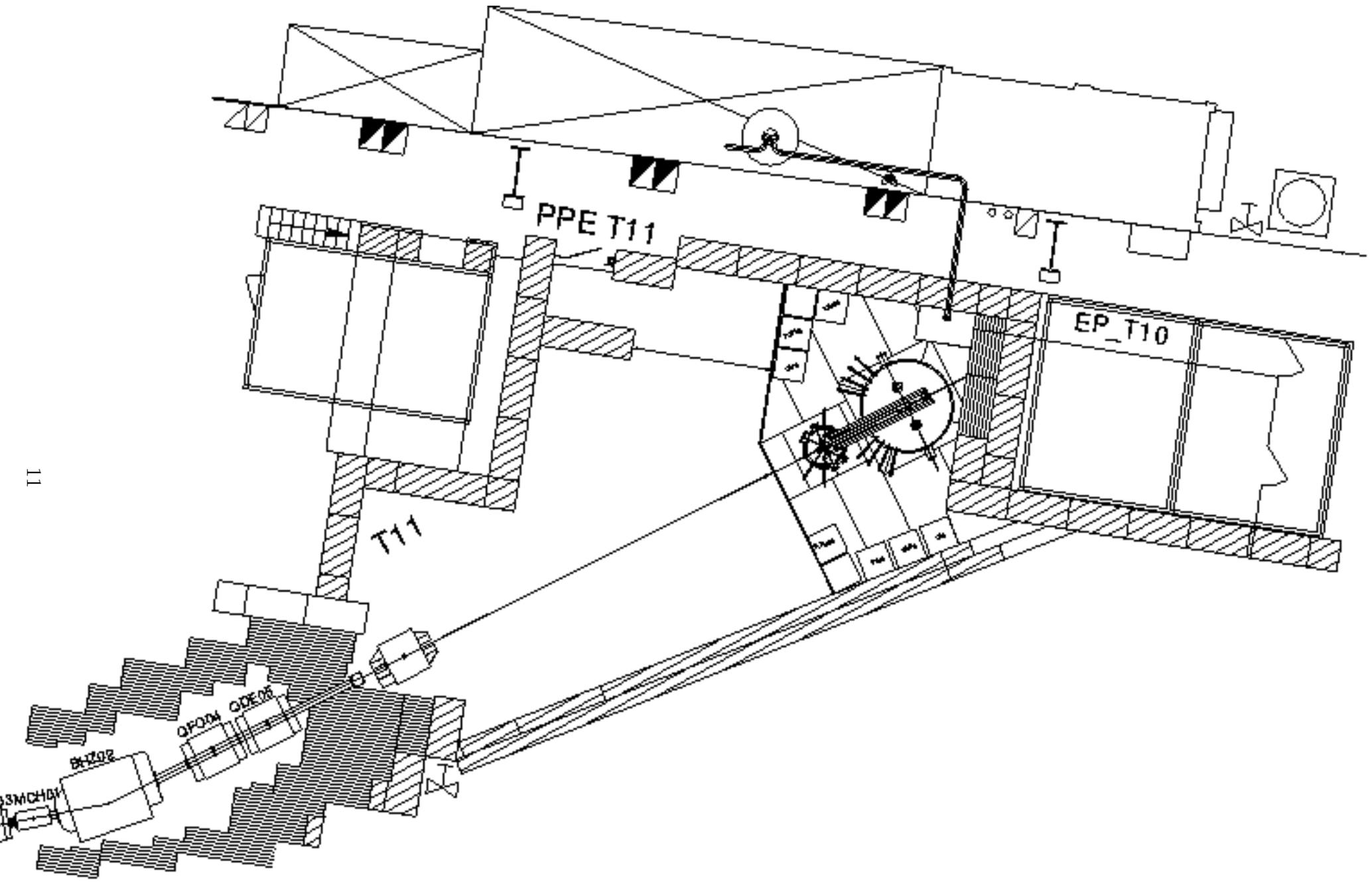,width=220mm}}
  \end{center}
  \caption{Updated experimental area layout of the CLOUD facility at
the CERN PS in the T11 beamline of the East Hall.}
  \label{fig_beam_layout}    
\end{sidewaysfigure}

\subsection{Experimental area layout}

\begin{figure}[tbp]  
  \begin{center}  
      \makebox{\epsfig{file=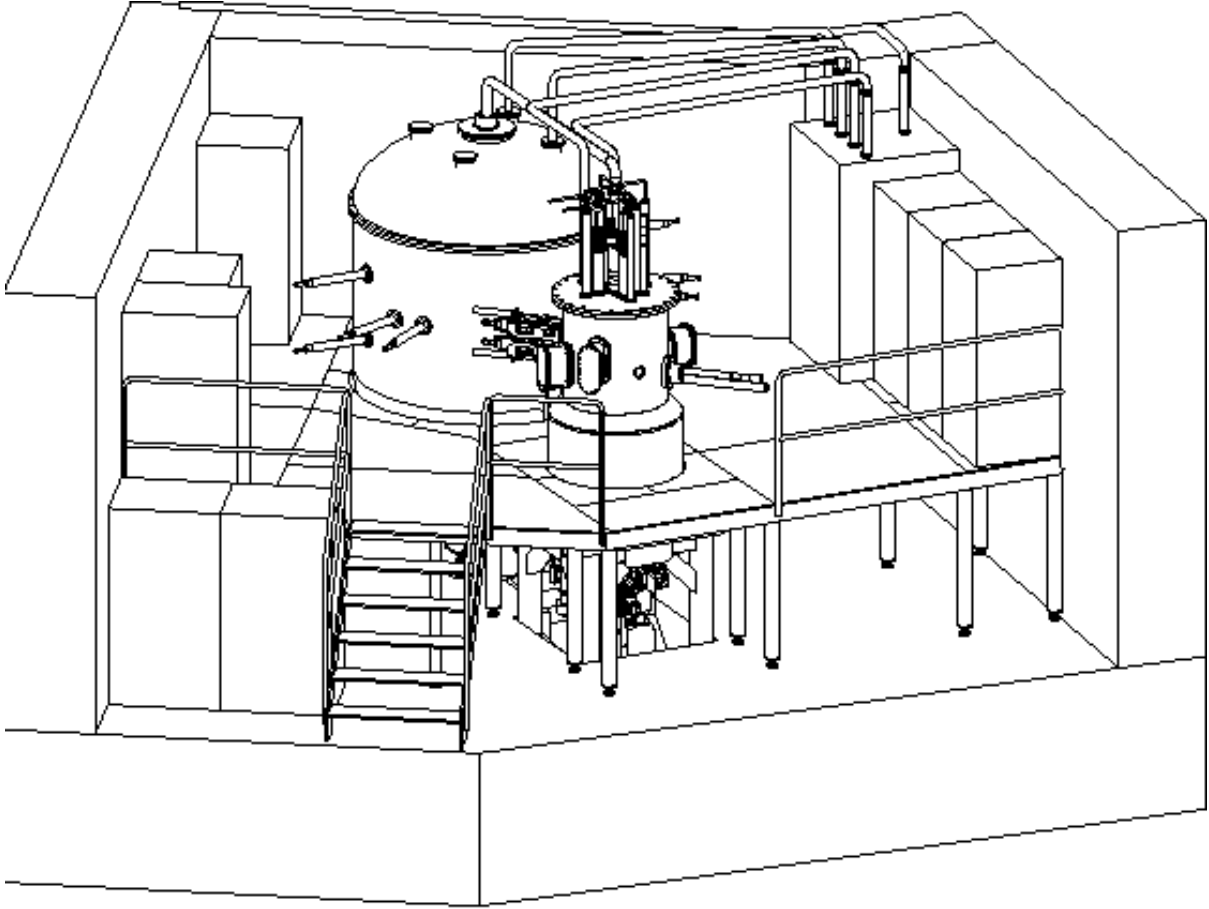,width=160mm}}
  \end{center}
  \caption{Perspective view of the CLOUD facility in the T11 beamline
of the East Hall}
  \label{fig_cloud_3d_east_hall}
\end{figure}

At its meeting on 6 September 2000, the SPSC  emphasized the importance
of maximising the remaining free space in the T11 beamline after CLOUD
is installed, in order to maintain the highest flexibility for future
detector tests.  Together with PS Division staff we have therefore
developed a new beam layout, which is shown in
Figs.\,\ref{fig_beam_layout} and \ref{fig_cloud_3d_east_hall}.  By
re-configuring the concrete shielding blocks at the end of the present
T11 beamline and by minimising the footprint of the facility, the free
space along the beamline will decrease by only 1 m after CLOUD is
installed (from the present 11 m to 10 m). 

\section{Background measurements in the East Hall}

Another question raised by the SPSC concerned the ambient background
level  in the East Hall.   Accordingly we have measured the atmospheric
small ion concentrations and variability in the East Hall during
operation of the CERN PS.

\subsection{Instrumentation}

The measurements were performed with a Programmable Ion Mobility
Spectrometer (PIMS) which we have developed in Reading  \cite{aplin00a}.
This device involves a sampling cylinder in which the ions are drifted
onto a well-insulated axial electrode under the influence of an
electric field.  The PIMS is ventilated by a fan which draws air
through the cylinder at about 2 ms$^{-1}$.  A sensitive current
amplifier  integrates the ion charges collected. The small ion
concentration is directly proportional to the air conductivity, which
is calculated from the measured current, and the sampling tube's
geometry and ventilation rate. Typical currents are a few 100
femtoamperes.  The PIMS is calibrated by means of two operating
modes---ion-induced voltage decay and direct ion current
measurement---under microprocessor control \cite{harrison01}.  A
rigorous calibration procedure has been developed for atmospheric
operation under demanding and variable environmental conditions
\cite{aplin00b}, including a new \textit{in situ} method for direct
calibration of the current amplifier \cite{harrison00}.  

\begin{figure}[htbp]
  \begin{center}
      \makebox{\epsfig{file=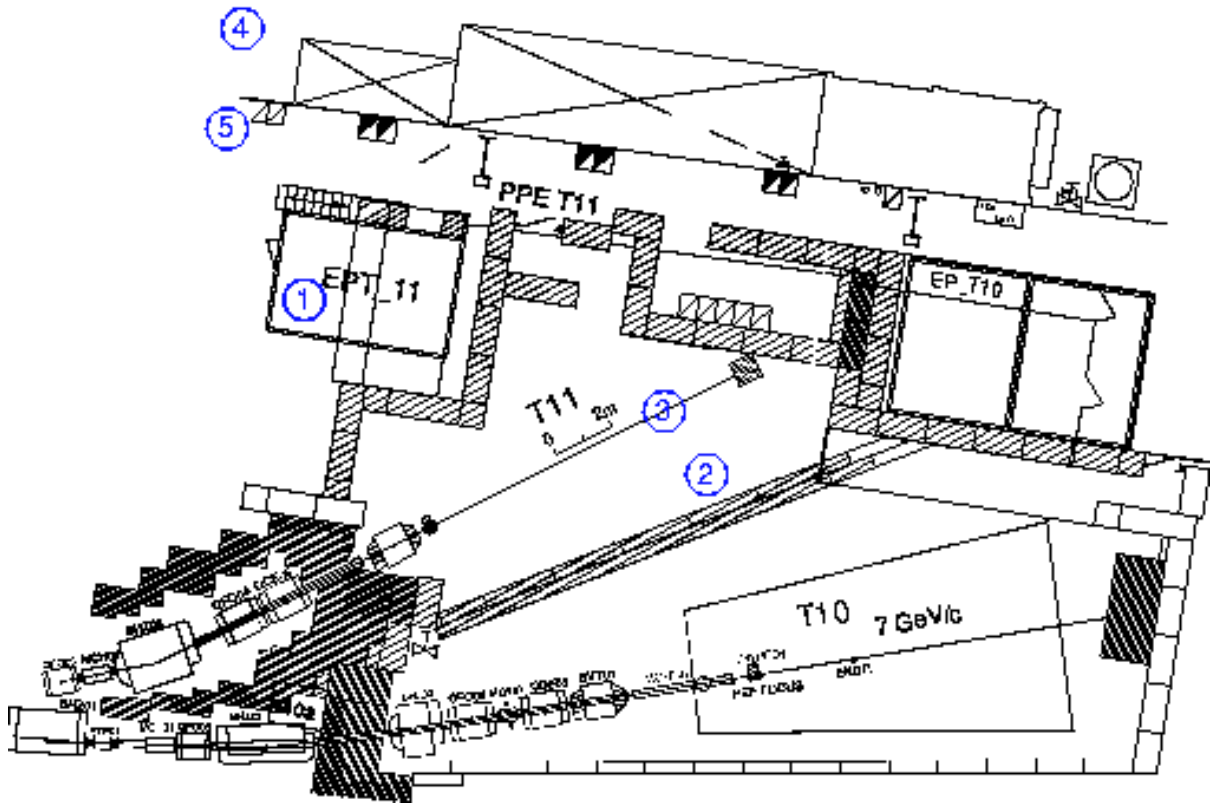,width=160mm}}
  \end{center}
  \caption{Locations of the five measurement sites in the CERN East
Hall.}
  \label{fig_ion_test_sites}    
\end{figure}

\subsection{East Hall tests}

Atmospheric small ion concentrations were measured at various East Hall
locations, as indicated in Fig.\,\ref{fig_ion_test_sites}, on 5th
October 2000 from 11h30-17h00 local time.  During this time the CERN PS
was delivering about 3 pulses per supercycle onto the East Hall Target
North (the target for the T9, T10 and T11 beamlines) at an average
intensity of $2 \times 10^{11}$ protons/pulse.  During the tests the T11
beam was off and the adjacent T10 beam was on. 

At each site, the PIMS instrument was operated using an automated
sequence of voltage decay measurements (sampled at 2 Hz) and current
measurements (the microprocessor calculates averages of ten 1 Hz
samples) every three minutes. The system leakage current and op-amp
input offset voltage were also sampled to compensate the measured ion
currents for temperature-dependent effects \cite{aplin00a}. 

Ion measurements were made for approximately 30 minutes at each
location.   The negative air conductivity and negative ion
concentration were calculated as described in
\cite{aplin00b}, and are presented in Table
\ref{tab_ion_measurements}. The calculation of ion concentration
assumes that the negative small ions have a mean mobility of 1.9
cm$^2$V$^{-1}$s$^{-1}$ \cite{mohnen}.

\begin{table}[htbp]
  \begin{center}
  \caption{Measurements of the atmospheric negative ion conductivity
and negative small ion number concentration in the East Hall. The
negative ion concentrations are approximately equal to the ion pair
concentrations. During all measurements the PS was operating and
delivering beam to Target North (the T9/10/11 target) and the adjacent
 secondary beam (T10) was on.}
  \label{tab_ion_measurements}
  \vspace{5mm}
  \begin{tabular}{| r | l | c c  | c c |}
  \hline
  \textbf{No.} 
 & \textbf{Location} & \multicolumn{2}{|c|}{\textbf{Conductivity}  } & 
  \multicolumn{2}{|c|}{\textbf{Ion concentration}  }  \\
    &  & \multicolumn{2}{|c|}{\textbf{(\boldmath $ \times 10^{-15} $ 
     \unboldmath S/m)}  } & 
  \multicolumn{2}{|c|}{\textbf{(cm\boldmath $^{-3}$ \unboldmath)} } 
\\
    \cline{3-6}  
  & & Mean & $\sigma$ & Mean & $\sigma$  \\
  \hline
  \hline
  &  &  & & &  \\[-2ex]
   1 & T11 control room & 12.2 & 5.6 & 401 & 184 \\
   2 & T11 beam area, 3m off-axis, beam off & 8.1 & 6.6 & 266 & 217 \\
   3 & T11 beam axis, beam off & 12.9 & 6.3 & 424 & 207 \\
   4 & Outside East Hall & 13.5 & 5.0 & 444 & 197 \\
   5 & Inside East Hall & 14.1 & 6.0 & 464 & 197 \\[0.5ex]
  \hline
  \end{tabular}
  \end{center}
\end{table}

Typical atmospheric air negative conductivities (which are directly
proportional to the small ion concentrations) in urban air at Reading
are around $13 \cdot 10^{-15}$ S/m, with short-term (of order tens of
minutes) variability up to 100\% (see, for example, Fig.\,37 on p.\,58
of the CLOUD proposal).  These conductivities correspond to around
400~ion~pairs~\pcc, and are comparable to ground-level values measured
elsewhere (e.g. Fig.\,\ref{fig_cosmics}b).  The values measured in the
East Hall are quite similar, and show no evidence for any significant
increase in the ambient ion background.  These measurements provide a
realistic test of backgrounds for CLOUD since air is the active
material in the detector and ion number concentration is the parameter
of interest. We conclude that low background conditions exist in the
East Hall for the CLOUD studies.

\section{Conclusions}

We are proposing a European facility at CERN where atmospheric
scientists can investigate the role of natural ionisation in aerosol
and cloud formation.  The concept of a facility is appropriate for the
comparatively large and complex experimental programme of CLOUD
extending over several years, and in view of the need for flexibility
in a field where rapid progress may be expected in the next few years.

CLOUD's requirements include a variable particle beam, techniques
derived from CERN's bubble-chamber experience, exacting cryogenic
temperature control, and the skills in integration, experimental
management and data-processing for which CERN is well known. For all
these reasons, CERN is uniquely suited to host the facility. 

The facility would use a 3.5 GeV/c secondary beam from the CERN PS in
the T11 beamline of the East Hall. Beam intensities between 100 and
$10^5$ particles/pulse are required to cover the desired range of 1--10
$\times$ the cosmic ray intensities found in the atmosphere.  We have
experimentally verified low background conditions in the East Hall for
the proposed studies.  

The CLOUD detector would occupy a permanent space at the end of the T11
beamline, reducing the free space available for test experiments from
the present 11 m to 10 m.  CLOUD would take data during about one half
of the yearly operation of the East Hall, leaving the T11 beamline
available for test experiments during the remaining time.

Unique in the world, this facility will open up an essentially new
field of atmospheric research. Its primary task, as described in the
CLOUD proposal, is to pursue the question of how cosmic rays may
influence cloud microphysics. If clouds respond to the solar variations
that modulate the cosmic rays reaching the Earth, there are 
consequences for the evaluation of climate change. To settle the issue,
one way or the other, is therefore of urgent global importance.

\section*{Acknowledgements}

We would like to thank Jean-Pierre Delahaye, Luc Durieu, Dieter
M\"{o}hl and Jean-Pierre Riunaud of the CERN PS Division for their
invaluable assistance.

\end{document}